\documentstyle[12pt]{article}
\def\tsetrue{T} \def\tsefalse{F} 

\let\tseepsfon\tsefalse    
\let\tsepaper=\tsefalse   
\let\tsenoteon=\tsefalse   
\let\tselse=\tsefalse     
\let\tseletter=\tsetrue  
\let\tsedevon=\tsefalse   

\newcommand{\tsecompldate}{26th July, 1993}
\newcommand{\tseprepno}{Imperial/TP/92-93/45} 
\newcommand{\tsehepphno}{hep-ph/9307335}

\if\tsetrue\tsedevon \newcommand{\tsedevelop}[1]{{#1}}
\else \newcommand{\tsedevelop}[1]{{}}
\fi

\if\tsetrue\tsepaper 
\else 
\fi

\tsedevelop{\typeout{*** T.S.E. Development mode on ***}}

\if\tsetrue\tseepsfon 
\else 
\fi

\if\tsetrue\tsepaper 
\else 
\fi

\if\tsetrue\tsepaper  
\fi

\if\tsetrue\tseletter{
\typeout{          as for Letter paper ---}
}\else
\fi
\if\tsetrue\tselse{\makeatletter

\@addtoreset{equation}{section}
\makeatother
\fi
\if\tsetrue\tselse
\typeout{--- Equations labeled as (section.equation) ---}
\fi

\newcommand{\vol}[1]{{\bf #1}}

\if \tsedevon\tsefalse \newcommand{\tselea}[1]{\label{#1}}
\else \newcommand{\tselea}[1]{\label{#1} \\
& \mbox{ } & \mbox{ (#1) } \nonumber }
\typeout{--- Renewing tselabel command ---}
\fi
\newcommand{\tseleq}[1]{\tsedevelop{\mbox{  (#1) }}\label{#1}}

\newcommand{\tbib}[1]{\bibitem{#1}\tsedevelop{ [#1] }}
\newcommand{\tref}[1]{(\ref{#1}\tsedevelop{-#1})}

\newcommand{\tcite}[1]{\cite{#1}\tsedevelop{ [#1] }}

\newcommand{\tnote}[1]{\if\tsetrue\tsenoteon \footnote{#1} \fi}
\if\tsetrue\tsenoteon{
\typeout{--- Tim Footnotes Included ---}
}\else \typeout{--- Tim Footnotes Excluded ---}
\fi

\newcommand{\tftitleone}{The bubble diagram.}
\newcommand{\tftitletwo}{The $\omega$ integrand of $B_c$  for the
different possible orders of the limits.}
\newcommand{\tftitlethree}{The high temperature results of $B_c$  for the
different possible orders of the limits.}

\if\tsetrue\tseepsfon \newcommand{\tseepsffile}[1]{\epsffile{#1}}
\newcommand{\tseepsfxsize}[1]{\epsfxsize=#1}
\else \newcommand{\tseepsffile}[1]{{}}
\newcommand{\tseepsfxsize}[1]{{}}
\fi

\newcommand{\tcaption}[2]{
\if\tsetrue\tsepaper \vspace{5cm} \caption{  }
\else
\vspace{#1} \caption{#2}
\fi }

\newcommand{\half}{\frac{1}{2}}
\newcommand{\bea}{\begin{eqnarray}}
\newcommand{\eea}{\end{eqnarray}}
\newcommand{\beq}{\begin{equation}}
\newcommand{\eeq}{\end{equation}}
\newcommand{\nnel}{\nonumber \\ {}}
\typeout{--- Equation break set for wide text ---}
\newcommand{\npagepub} {\if\tsetrue\tsepaper{\newpage }\fi}
\if\tsetrue\tsepaper{
\typeout{--- Page Breaks for Pub. Version ON ---} }\else{
\typeout{--- Page Breaks for Pub. Version OFF ---} }\fi

\newcommand{\ra}{\rightarrow}

\newcommand{\dt}[1]{(2\pi)^{-3}\int d^3\vec{#1}}

\newcommand{\dtk}{\dt{k}}

\newcommand{\veck}{\vec{k}}
\newcommand{\vecp}{\vec{p}}
\newcommand{\vecx}{\vec{x}}
\newcommand{\vecv}{\vec{v}}

\newcommand{\texpect}[1]{\mbox{ Tr} \{e^{-\beta H} #1 \}
/\mbox{Tr} \{ e^{-\beta H} \} }

\newcommand{\real}[1]{{\rm Re} \{ #1 \}}
\newcommand{\imag}[1]{{\rm Im} \{ #1 \}}

\begin{document}

\typeout{--- Title page start ---}

\if\tsefalse\tsepaper \thispagestyle{empty}\fi

\renewcommand{\thefootnote}{\fnsymbol{footnote}}

\begin{tabbing}
\hskip 11.5 cm \= \tseprepno \\
\> \tsehepphno \\
\> \tsecompldate \\
\tsedevelop{\> (\LaTeX -ed on \today ) \\}
\end{tabbing}
\vskip 1cm

\begin{center}
{\Large\bf Thermal Bubble Diagrams} \\
{\Large\bf Near Zero Energy}
\vskip 1.2cm
{\large\bf T.S. Evans\footnote{E-mail: T.Evans@IC.AC.UK}}\\
Blackett Laboratory, Imperial College, Prince Consort Road,\\
London SW7 2BZ  U.K.
\end{center}
\if\tsetrue\tsepaper{\begin{center}
Tel: U.K.-71-589-5111 ext. 6980 \\
Fax: U.K.-71-589-9463 \\
\mbox{  }
\end{center}
}\fi

\npagepub
\vskip 1cm
\begin{center}
{\large\bf Abstract}
\end{center}


The zero four-momentum and equal mass limits are taken for the
bubble diagram of scalar fields.  It is seen that RTF and ITF are in
complete agreement.  However contributions from this diagram to both
retarded and time-ordered functions do depend on the order of the
limits and can be infinite in some cases.  This shows explicitly
that the relation between the free energy and a derivative expansion
of a thermal effective action is generally much more complicated
that is the case at zero temperature.

\vskip 1cm

\renewcommand{\thefootnote}{\arabic{footnote}}
\setcounter{footnote}{0}

\npagepub

\typeout{--- Main Text Start ---}

The simplest example in relativistic field theory of a diagram with
non-trivial momentum dependence is the bubble diagram of figure
\tref{fbub}.  \if\tsepaper\tsefalse
\typeout{figure: BUBble diagram}
\begin{figure}[htb]
\tseepsffile{bub.eps}
\caption{\tftitleone}
\label{fbub}
\end{figure}
\fi It
is therefore illuminating to see exactly what results the standard
Feynman rules lead to especially as they turn out to be non-trivial
in the zero four-momentum, equal mass limit.
This limit is also of special interest because the zero energy limit is
closely tied to the infra-red behaviour which plays a vital role in
both first and second order phase transitions as, for example,
the discussion of
the cubic term in the free energy for electroweak theory shows
(see \tcite{pt} and references therein).

However the zero energy limit is problematical.   In the context of
the calculations done using the Feynman rules of RTF (Real-Time
Formalisms) \tcite{Raybook,TFD,NS1,NS2,LvW,TSEnrtf}
pathological terms of the form $[
\delta(K^2-m^2) ]^{N\geq 2}$ appear
\tcite{LvW,MOU,FMUO,FGnO,TSEbub,TSEzm,TSEze,TSEwpg}. One of the
motivations of this paper is to show that the obvious problems of
RTF at zero energy do not indicate that RTF is flawed but rather
that thermal Green functions at zero energy are intrinsically more
complicated than one would expect from experience with zero
temperature field theory.  Further, using the `simple'  scalar bubble
diagram as an example, it will be clear that both RTF and ITF
(Imaginary-Time Formalism) \tcite{LvW,ITF} are in complete agreement
and show the same difficulties with the zero energy limit.
It is hoped that this example will clarify the nature of
the calculational problems with this limit.  In doing so we will be
extending existing calculations of the bubble diagram,
\tcite{TSEbub,TSEzm,AVBD}, to cover all possible variations.

We start by noting the essential properties of two-point Green
functions of equilibrium field theory as well as their relation to
the results of RTF or ITF calculations.  This will also serve to
establish the notation used here.

In RTF one has a direct handle on time-ordered Green functions.  The
connected two-point function in RTF, $ \Pi^{ab}$, is a two by two
matrix where by definition the $11$ component is the real-time
time-ordered function, $\Pi_t$
\beq
 i  \Pi^{11}(t,\vecx) = i \Pi_t(t,\vecx)
= \texpect{\mbox{T} \phi(t,\vecx) \phi(0,0) }
\tseleq{e11t}
\eeq
This is true whatever version of RTF is used ($C^\ast$-algebra,
Thermo Field Dynamics, path integral methods, etc.).

In ITF one is initially calculating the Euclidean time-ordered
function, $\Pi_I$.  In terms of energy it is only known at
imaginary energies which are
multiples of $2 \pi i / \beta$.  Only by making an analytic continuation
can one look at the the behaviour of ITF calculations in the
neighbourhood of zero
energy.  Using standard boundary conditions at infinite energies
this can be done \tcite{BM,TSEnpt} and the resulting function will
be denoted by $\Pi_c$.
One finds that at real energies ITF is
calculating the retarded ($R$) and advanced ($A$) Green functions
\bea
\Pi_c(z=E+ i  \epsilon) &=& R(E) \nnel
\Pi_c(z=E- i  \epsilon) &=& A(E)
\tselea{ecra}
\eea
where $E$ will be used to indicate real energies.\tnote{The
three-momenta or space arguments are not always given when this
dependence is straightforward.}  As functions of time,
the retarded and advanced functions are given by
\bea
iR(t,\vecx) &=&  \theta( t) \texpect{[\phi(t,\vecx) \phi^\dagger (0,0)]}
\\
iA(t,\vecx) &=& -\theta(-t) \texpect{[\phi(t,\vecx) \phi^\dagger (0,0)]}
\eea
where only bosonic fields are being considered.

Standard relations between the
time-ordered and retarded functions are simple to derive \tcite{LvW} and
this then provides a link
between the usual results of RTF and ITF calculations by using
\tref{e11t} and \tref{ecra}.
In particular relations can be found for truncated diagrams
\tcite{LvW,TSEwpg}
\tnote{Also $\Pi^{-1}(E)]^{11} = [1+ n(E)] R(E)^{-1}  \pm
n(E) A(E)^{-1}$}
\bea
\imag{[\Pi^{-1}(E)]^{11}  } &=& \frac{e^{\beta E} +1}{e^{\beta E} -1}
\imag{ R(E) }
\tselea{epiinvim},
\\
\real{[\Pi^{-1}(E)]^{11}  } &=& \real{R(E)}
\tselea{ertre}
\\
\imag{ R(E=0)^{-1} }  &=& 0
\tselea{erim}
\eea
where
\bea
n(E) &=& \frac{1}{e^{\beta E} - 1}
\tselea{endef}
\eea
Such relations are derived from the definitions of these full Green
functions but they   hold even when the contribution from only one
single Feynman diagram is considered.  This is not surprising  as the
relations come directly from the definitions of what is meant by
equilibrium thermal Green functions (cyclicity of the thermal
trace, Kubo-Martin Schwinger condition etc.) which must be respected
order by
order if an approximation is to be physically realistic.

The usual problem of the infra-red divergence in the Bose-Einstein
distribution seems to hit us straight away and we find
 for the bosonic case
\bea
\lim_{E\rightarrow 0} \{ [\Pi^{-1}(E)]^{11} - R(E)^{-1} \}
&=& \lim_{E\rightarrow 0}  \frac{2}{\beta E} \imag{ R(E)^{-1} }
\tselea{etrtr}
\eea
The divergence in \tref{etrtr} looks very suspicious
in view of the RTF divergences we will discuss below.
However \tref{etrtr} was derived for arbitrary $\vecp, m_1,m_2$
whereas it is the behaviour at equal mass and zero four-momentum
that will interest us below.   Further, as noted above,
$\imag{R(E=0)}=0$ and
a more careful analysis \tcite{TSEwpg} shows that
\bea
\lim_{E\rightarrow 0} \{ [\Pi^{-1}(E)]^{11} - R(E)^{-1}  \}
&=& \int dt \; \texpect{\phi_\mu(t,\vecx) \phi_\nu^\dagger(0,\vec{0})}.
\tselea{etrtr2}
\eea
so that the obvious divergence in \tref{etrtr} has gone and the
difference may well be finite.  Thus \tref{etrtr} does
not seem to tell us much about possible divergences.

The problems which we wish to address here are most evident in the
RTF calculations where it is well known that singular terms do
appear.   These are the $[\delta(K^2-m^2)]^{N\geq 2}$ terms which
appear when $N$ lines in a diagram carry the same four-momentum and
are associated with particles of the same mass
\tcite{LvW,MOU,FMUO,FGnO,TSEbub,TSEzm,TSEze,TSEwpg}.  This occurs
when parts of a diagram are self-energy insertions but in this case
it is simple to see that when the usual RTF sum over internal vertex
labelings is performed, such contributions cancel
\tcite{NS2,LvW,MOU,FMUO}.

However such singular terms also arise in diagrams where there are
external legs carrying zero four-momenta \tcite{FGnO,TSEzm}.  Such
diagrams are physically important,  for instance the free energy or
effective potential is also  the generating functional of all 1PI
diagrams in which any external legs carry zero four-momenta, though
this is not so straightforward for the thermal case
\tcite{TSEwpg}.\tnote{
One has to be rather more careful at non-zero temperature.  It has
been known for sometime that the free energy was {\it not} given in
terms of zero four-momentum 1PI RTF diagrams where all the external
legs were fixed to be type 1.  These diagrams are just the
time-ordered thermal Green functions.   Instead as was  shown in
\tcite{LvW,MOU,FMUO,FGnO,TSEzm} one must keep just one vertex fixed
to be type one and all others summed over.  By using the results of
\tcite{TSEnpt} in the zero energy limit one can see \tcite{TSEwpg}
that this is the prescription for getting the zero energy
retarded/advanced functions (they are all identical at zero energy)
out of RTF diagrams.  Thus the free energy is only known to be a
generating functional of zero four-momenta 1PI thermal retarded
Green functions.}


The simplest example of the $[\delta(K^2-m^2)]^{N\geq 2}$
singularities comes from the bubble self-energy diagram \tref{fbub}
where two scalar
fields run round the loop.  The contribution to the time-ordered
function, $B_t$, is easily calculated in RTF and is given by
\bea
-i B_t(E,p;m_1, m_2) &=& - i  B^{11} \nnel
&=&  \frac{(- i  g)^2}{2}  \int \frac{d^4K}{(2\pi)^4}
i \Delta^{11}(K,m_1)  i  \Delta^{11}(K+P,m_2)
\nnel
&=&  - i  B_{t0} - i  B_{t1}- i  B_{t2}
\nnel
\eea
where
\beq
i \Delta^{11}(K,m) = \frac{i}{K^2-m^2 + i \epsilon} + n(|k_0|)
\delta(K^2-m^2)
\eeq
It is convenient to split it into three terms, $B_{tj}$ being the
contribution to the time-ordered functions from the terms in the RTF
calculation containing $j$ delta functions and Bose-Einstein factors.

Throughout $E$ will represent real external energies with
$P^\mu=(E,\vecp)^\mu$ the external Minkowskii four-momentum,
while $K^\mu=(k^0,\veck)^\mu$
denotes Minkowskii loop momentum.  The modulus of
three-momentum is denoted by $p=|\vecp|,k=|\veck|$.

The $B_{t2}$ contains a product of two delta functions.
This is only a problem
for the RTF calculation if we have set $m_1=m_2$ and $E,p=0$ before
doing the integrals.
Thus avoiding any one of these limits is sufficient to give a well
defined integral and this is what will be done.
Doing the energy and angular integrals
gives\tnote{TSE notes `RTF Scalar Bubble Im Parts II'}
\bea
B_{t2}(E,p;m_1, m_2)  &=& \frac{- i  g^2}{16 \pi p} \sum_{\pm} \int dk
\frac{k}{\omega_1} n(\omega_1) n(\Omega_2) \theta(\cos \theta^{\ast} -1)
|_{\theta = \theta^\ast}
+ (1 \leftrightarrow 2)
\tselea{et2}
\eea
where
\bea
\cos \theta^{\ast} &=& \frac{1}{2 p k }
(P^2 + m_1^2 - m_2^2 \pm 2 E \omega_1)
\\
\cos(\theta) &=& \frac{\vecp . \veck }{ p k}
\eea
The dispersion relations, $\omega_i$, $\Omega_i$ are defined to be
\beq
\omega_i = k^2 + m_i^2, \; \; \;
\Omega_i = (\vecp+\veck)^2 + m_i^2   ,
\eeq
Note that in general the two delta functions can both be non-zero .
Thus $B_{t2}$ is often non-zero and well behaved as \tref{et2} shows.

If we keep $m_1\neq m_2$ then as $E,p \rightarrow 0$ we find
\bea
\lim_{E,p \rightarrow 0} B_{t2}(E,p;m_1 \neq m_2) &=& 0
\eea
as the argument of the two delta functions can not be satisfied
simultaneously.
However setting $m=m_1=m_2$ first gives an interesting result.  Taking
the limit $E,p \rightarrow 0$ the ratio $v=|\vecv|$ can be kept fixed,
where
\beq
\vecv = \frac{\vecp}{E} \; \; , \; \; \gamma = (1-v^2)^{-\half}.
\tseleq{evdef}
\eeq
We then find
\bea
\lim_{E,p \rightarrow 0} \lim_{ m_1 \rightarrow m_2}
B_{t2} (E,p;m_1 \neq m_2)
&=&
\frac{- i  g^2}{8 \pi} \frac{m x_c}{|p|} \theta(v^2-1)
\int_1^\infty dx \; n^2(mx_cx)  .
\tselea{et2emv}
\\
&\simeq &
\frac{- i  g^2}{8 \pi} \theta(v^2-1)
\left[ \frac{T^2}{m x_c |p|} + \right. \nnel
&& \left.
\frac{T}{|p|}(-\half -\zeta_0(1) - \psi(1) +\log (b) )
\right]
\eea
where
\beq
x_c = | v \gamma | .
\eeq
Throughout we shall use $\simeq$ to indicate that a high temperature
expansion, $ T \gg E,p,m_1,m_2$ has been taken but only terms of $O(T)$
and bigger are given.
The integral is generally finite but does not have a simple form so
for illustrative purposes the high temperature limit,
$T \gg m_1,m_2 \gg E,p$, is given.

In taking the small $E,p, (m_1-m_2)$ limits,  $B_{t2}$, is generally
well defined except at the $E,p=0$ $m_1=m_2$ point itself where
$B_{t2}$ is infinity or zero.  It is also pure imaginary.  However
the residue of the pole depends on how this point is approached in
the $E,p$ plane.  So the delta-squared term of the RTF calculation
does reflect genuine infinites in the $B_{t2}$ term.

As $B_{t2}$ is imaginary, it is useful to look at the remaining
temperature dependent imaginary parts coming from $B_{t1}$ in case
it cancels any of the odd behaviour in $B_{t2}$.  While $B_{t1}$
does not contain any explicit $[\delta(K^2-m^2)]^2$ terms, which
provided some of the original motivation for studying $B_{t2}$, we
find
\bea
B_{t1}(E,p;m_1,m_2) &=& \frac{g^2}{2} \sum_{\pm} \dtk \;
\frac{n(\omega_1)}{2 \omega_1}
\frac{1}{(E \pm \omega_1 )^2 - \Omega^2_2+ i \epsilon}
\nnel && +
\frac{n(\Omega_2)}{2 \Omega_2}
\frac{1}{(E \pm \Omega_2 )^2 - \omega^2_1 + i \epsilon}
\tselea{ebt1dtk}
\eea
from which
\bea
\imag{B_{t1}} &=&
\frac{- i  g^2}{16 \pi p} \sum_{\pm} \int dk
\frac{k}{\omega_1} \half \left( n(\omega_1) + n(\Omega_2) \right)
\theta(\cos \theta^{\ast} -1) |_{\theta = \theta^\ast}
\tselea{et1im}
\eea
Results are similar to $B_{t2}$.  First
\bea
\lim_{E,p \rightarrow 0} \imag{ B_{t1} (E,p;m_1 \neq m_2)} &=& 0    .
\eea
Then we have in the high temperature limit $T \gg m_1,m_2 \gg E,p$
\bea
\lim_{E,p \rightarrow 0} \lim_{ m_1 \rightarrow m_2} \imag{ B_{t1} } &=&
\frac{- i  g^2}{8 \pi} \theta(v^2-1)
\frac{T}{|p|}
\eea
In fact it is easy to calculate the exact total temperature dependent imaginary
part and
this is
\bea
\lim_{E,p \rightarrow 0} \lim_{ m_1 \rightarrow m_2} \imag{ B_{t1+2} } &=&
\frac{- i  g^2}{8 \pi} \theta(v^2-1) \frac{T}{|p|} n(\beta m x_c)
\tselea{eimt}
\\
&\simeq & \frac{- i  g^2}{8 \pi} \theta(v^2-1)
\left[ \frac{T^2}{m x_c |p|} - \frac{T}{2|p|} \right] \nnel
x_c &=& | v \gamma|
\tselea{exc}
\eea
Both $B_{t1}$ and $B_{t2}$ contribute to this divergent result
despite the fact that $B_{t1}$ had only a single delta function in
its integrand.  However, away from
$m_1=m_2$  and $E,p\rightarrow 0$ the result is finite.  For
instance taking the $v=\infty$ limit in \tref{eimt} ( i.e. $p$
small but not zero) shows that its divergence only occurs at the zero
four-momentum point and is not associated with just the zero energy
limit as \tref{etrtr} might suggest.  So the delta-squared
term in $B_{t2}$ is therefore flagging a deeper problem,
the existence of genuine divergences
present in the total result for the imaginary part of the
time-ordered function,
$\imag{B_{t}}$, at the zero four-momentum equal mass point.

Having established this situation with the imaginary part of the
time-ordered result as obtained in an RTF calculation, it is
interesting to see what is happening with other results obtained from
the same diagram at the same point.  So we will look at the real
part of the time-ordered function obtained in RTF, at the ITF
results from which we will extract the advanced and retarded Green
functions, and check the raltion between them.
This will enable us to see if RTF calculations or
time-ordered functions are unreliable in this limit and whether ITF
calculations or retarded Green functions have similar problems.

For the specific case of the
bubble diagram in ITF one initially calculates
$B_I(2 \pi i n / \beta,p;m_1,m_2)$ which
is only defined at discrete values of the energy,
$2 \pi  i  n / \beta$, $n \in {\cal Z}$.  The only real value of
energy which is directly accessible is $E=0$.
In particular, it should be noted that in
the ITF calculation setting $m_1=m_2, E,p=0$ before integrating
does gives a unique and well defined result.  There is no divergence in
the integrand as there was with RTF and this result will be
noted below, \tref{eep}.  It in the other ITF calculations discussed
here, the $m_1=m_2, E,p=0$
limit is taken after the loop energy integration.

The ITF function, $B_I$,
can be continued to one defined at general complex energies
which is denoted by $B_c$.  It is found to be
\bea
\lefteqn{B_c(z\in {\cal C},p;m_1,m_2) = (T=0)\; \; + }
\nonumber \\
&&\frac{g^2}{2} \sum_{\pm} \dtk \left[
\frac{n(\omega_1)}{2 \omega_1}
\frac{1}{(z \pm \omega_1)^2 - \Omega^2_2}
\; \; + \frac{n(\Omega_2)}{2 \Omega_2}
\frac{1}{(z \pm \Omega_2)^2 - \omega^2_1} \right]
\tselea{ebcdtk}
\eea
where $z$ is the complex external energy parameter.
The retarded, $R(E,p;m_1,m_2)$, and advanced, $A(E,p;m_1,m_2)$, functions
correspond to continuations of $B_c$
to either side of the real energy axis,
$z \rightarrow E \pm  i  \epsilon$ \tcite{BM,TSEnpt}.
By inspection the $ i
\epsilon$ terms are different from the RTF $B^{11}$ calculation in
which the energy integral has been performed.  This allows the
imaginary part of the ITF calculation to differ from the RTF result.
 We find for general $p,m_1,m_2$ and zero energy that
\bea
B_c(z = 0 +  i  \epsilon,p;m_1,m_2) &=& R(E=0,p;m_1,m_2)
= A(E=0,p;m_1,m_2)
\nnel
&=&
B_I(z=0 ,p;m_1,m_2) \nnel
&=&\real{ B^{11}(E=0,p;m_1,m_2) }   .
\eea
The last result is obtained by inspection.  These results
are consistent with the identities
\tref{ertre}, \tref{erim} and \tref{etrtr}.  Note that
not surprisingly the results obtained using $B_I$
at the zero energy point
are identical with the case
where an analytic continuation is made to general complex energies and
then $E\rightarrow 0$ is taken before any other limit.
However, these results are questionable if the functions
are badly behaved at any point.  As we have already seen problems
when $E,p=0, m_1=m_2$, it remains to be shown that the above
relations hold at that point.

The next stage is to look at the $E,p=0, m_1=m_2$ limit of the ITF
calculation of $B_c$.  The limit can be taken in various ways and it
is easily found that\tnote{see ``ITF scalar bubble'' notes pp3 (30-8-92),
pp4 (8-1-92), and end for High temp.}
\bea
\lim_{ m_1 \rightarrow m_2} \lim_{E,p \rightarrow 0}
B_c(E,p;m_1,m_2) &=& (T=0) +
\frac{ g^2}{4 \pi^2} \frac{d}{dm^2}
\int_0^\infty dk \; \frac{k^2}{\omega} n(\omega) \nnel
&=& (T=0) - \frac{ g^2}{8 \pi^2}
\int_m^\infty d\omega \; \frac{1}{k} n(\omega)
\tselea{ebvinf}
\\
&\simeq&  - \frac{ g^2}{16 \pi} \frac{T}{m}
\tselea{ebvinfhit}
\\
&=& \lim_{p \rightarrow 0}
\lim_{ m_1 \rightarrow m_2} \lim_{E \rightarrow 0}
B_c(E,p;m_1,m_2)
\\
&=& \lim_{p \rightarrow 0}
\lim_{E \rightarrow 0} \lim_{ m_1 \rightarrow m_2}
B_c(E,p;m_1,m_2)
\nnel
&=& \lim_{p,m_1 \rightarrow m_2 } B_I(z=0,p;m_1,m_2)
\tselea{ebi}
\nnel
&=& B_I(z=0,p=0;m_1=m_2)
\tselea{eep}
\eea
Note that these $B_c$ results coincide with those from the direct,
no analytic continuation ITF calculation, i.e. $B_I$ \tref{ebi},
however the equal mass and zero three-momentum limits are taken.
In particular, it agrees with the result obtained when setting
$m_1=m_2$ and $E=p=0$ before
doing the energy sum which we have denoted as \tref{eep} with no
limit symbols.
The only one that causes problems is when $m_1=m_2$ is taken first
and then we take $E\rightarrow 0$ with or after $\vec{p} \rightarrow
0$.  If we are to get a single
answer we must take $E$ to zero before the energy integral is done or
before both of the other limits have been taken.  If this is not
done then more complicated
results are obtained and I find \tcite{TSEbub,TSEzm}
\bea
\lefteqn{\lim_{p, E \rightarrow 0, p/E=v} \lim_{ m_1 \rightarrow m_2}
B_c(E,p;m_1,m_2) }
\nnel
&=&
(T=0) - \frac{g^2}{8 \pi^2} \int_1^{\infty} d\omega \; n(\beta \omega)
\frac{ k}{\omega^2 + v^2\gamma^2 m^2}
\tselea{ebv}
\\
&\simeq& - \frac{g^2}{16 \pi} \frac{1}{1+\gamma} \frac{T}{m}
\tselea{ebvhit}
\eea
In the case $v=\infty$ ($E \rightarrow 0$ before $\vec{p} \rightarrow
0$) \tref{eep} is recovered.

This general result can be compared against calculations where
either $E$ or $p$ are kept non-zero and not small but in which, apart
from this, the usual limit is taken.  These calculations
give\tnote{see ``ITF scalar bubble'' notes pp3(30-8-92)}
\bea
B_c(E,p=0,m_1=m_2) &=& (T=0) - \frac{g^2}{8 \pi^2}
\int_1^{\infty}
d\omega \; n(\beta \omega) \frac{ k}{\omega^2 - E^2/4}
\tselea{epzEf}
\\
B_r(E=0,p,m_1=m_2) &=& (T=0) - \frac{g^2}{16 \pi^2} \frac{m}{p}
\int_1^{\infty} d\omega \; n(\beta \omega)
  \log \left[ \left(
\frac{ p+2k}{p-2k} \right)^2 \right]
\tselea{eEzpf}
\eea
In the zero four-momentum
limit the above results do indeed reproduce \tref{ebv} for $v=0$ and
$v=\infty$ respectively.\tnote{The $v=0$ high temperature limit is
$-\frac{g^2}{32\pi}\frac{T}{m} $ for comparsion with \tref{ebvinfhit}.}

We can also look at the imaginary part of $B_c$ in the case where
$m_1=m_2$ is taken first \tref{ebv}.
This expression for $B_c$ is real for $|v| <1$ but for $|v|>1$ the
integrand has
a pole and an imaginary part is generated.  The small $ i \epsilon$
present in the energy, and hence in $v$,
tells us which way round the pole to go and we find\tnote{see
``ITF scalar bubble'' notes pp1(30-7-92)}
\bea
\lim_{p, E \rightarrow 0, p/E=v} \lim_{ m_1 \rightarrow m_2}
\imag{B_c(z=E+i\eta |\epsilon|,p , m_1,m_2)} &=&
\frac{ i g^2 \eta}{16 \pi}
\theta(v^2-1) n(\beta m x_c) \frac{E}{p}
\tselea{eimc}
\eea
Note that this ITF calculation of the bubble diagram's contribution
to the retarded function in the zero
four-momentum and equal mass limit is not divergent unlike the
the time ordered function.  However it is not generally zero as
\tref{erim} suggests.  While it is linear in energy near zero
energy, as the imaginary
part of the retarded function should be \tcite{TSEwpg}, there is a
competing $1/p$ factor as $p\ra0$ which keeps the result finite.
  It is only when the $v \rightarrow \infty$ limit
of \tref{eimc} is taken do we get zero for the imaginary part of
$B_r$ and so agreement with the spectral
representation result \tref{erim}.  This is not surprising as
\tref{erim} was derived for fixed $p$ and $E=0$ which corresponds to the
$v=\infty$ limit.

One can then use the spectral function results \tref{epiinvim}
together with these
ITF calculations of $B_c$ to tell us how
the imaginary part of the time-ordered function behaves.
It is completely consistent with the direct RTF calculation of the
time ordered function \tref{eimt}.  In particular the time-ordered
bubble diagram is found to be infinite for $v>1$.  However
it is not the factor of $1/E$ for small energies in
\tref{etrtr} that is responsible for the divergence, this is
canceled by the retarded result being linear in $E$.  Rather it is
the $1/p$ behaviour at small $p$ that is leading to the
singularities in the time-ordered function.

By inspection of \tref{ebt1dtk} and \tref{ebcdtk}, which differ only
in the $i\epsilon$ factors,
the real parts of the time-ordered function calculated using
RTF, $B^{11}$, and the retarded function calculated in ITF from
$B_c$, are seen to be equal even in the tricky limit.
Thus it is clear that
RTF and ITF are in complete agreement with each other.  If one choose
to, one could calculate the retarded function in RTF or the time
ordered function using ITF for the same diagram, and the same
results would be obtained.

On the other hand if we look at a different approaches to the
zero four-momentum point, $v$ finite, then we see that the retarded
function has a non-zero, finite value but the old infra red divergence
in the Bose-Einstein function ensures that the imaginary part of the
time-ordered function is blowing up in the region $1 < |v| < \infty$.
It is also zero for time-like limits $v<1$.

There is one approach to the $E,p,(m_1-m_2) \ra 0$ limit that has
not been considered so far and that is taking $p \ra 0$ first and
then taking the remaining limits in either order or simultaneously.
The result is found to be
\bea
\lefteqn{ \lim_{ E, (m_1 \rightarrow m_2) \ra 0} \lim_{p \rightarrow
0}
B_c(E,p;m_1,m_2) } \nnel
&=& (T=0) -
\frac{ g^2}{8 \pi^2} \int_0^\infty dk \left\{ \frac{k^2}{\omega^3}
n(\omega) +
\right.
\nnel
&& \; \; \; \; \; \left.
q^2 \beta m^2 \; n(\omega)[1+n(\omega)]
\frac{k^2}{\omega^2[q^2 m^2-\omega^2]}
\right\}
\nnel
&=& (T=0) -
\frac{g^2}{8 \pi^2} \int_m^\infty d\omega \; n(\omega) \left\{
\frac{k}{\omega^2} +
\right.
\nnel
&& \; \; \; \; \; \left.
 q^2 m^2
\frac{m^4 q^2+2\omega^4-3\omega^2 m^2}{k \omega^2[m^2 q^2-\omega^2]^2}
\right\}
\tselea{ebq}
\\
&\simeq& - \frac{g^2}{16 \pi} \frac{T}{m} \left[
1-\frac{1}{1+(1-q^2/4)^{1/2}} \right]
\tselea{ebqhit}
\eea
where
\beq
q=\frac{2(m_1-m_2)}{E}   .
\eeq
The $q=0$ limit means that $E \ra 0$ only after the other two and
gives the same answer as the $v=0$ limit of \tref{ebv}.
This shows that the
relative order
of the $m_1 \ra m_2$ and $p \ra 0$ is irrelevant, it only matters
how they are taken relative to $E \ra 0$ as has be found elsewhere.
The $q=\infty$ limit is
the same as \tref{ebvinf} as it should be.

Lastly we can take the high temperature limit first which allows us
to keep $E,p, m_1,m_2$ arbitrary relative to one another and yet get
a closed expression for the integral.
The leading term is $O(T)$ and is given by making the replacement
$n(\omega) \ra T/\omega$ (e.g. see appendix of \tcite{CCER}).
Thus by using
\tcite{GR}\tnote{ GR (4.296.4), checked numerically using MAPLE V.}
\bea
\lefteqn{ \int_0^\infty dy \frac{y}{y^2+b^2} \log \left[
\frac{y^2 + 2 a y \cos(t) + a^2}{y^2 - 2 a y \cos(t) + a^2} \right]
} \nnel
&=&
\half \pi^2 - \pi t + \pi \arctan \left[
\frac{(a^2-b^2)\cos(t)}{(a^2+b^2)\sin(t) + 2ab}\right]
\tselea{ebhit}
\\
&& a,b>0, \; \; \; 0<t<\pi \nonumber
\eea
we find that\tnote{I used $Y_{12} = 1+ \frac{m_1^2-m_2^2}{P^2}$}
\bea
B_c(E,p,m_1,m_2) &\simeq& - \frac{g^2}{32 \pi} \frac{T}{p}
\arctan \left[ \frac{2B(c+1)}{(c+1)^2 - B^2} \right] \; \; + (1
\leftrightarrow 2)
\\
B&=&\half \frac{p}{m_1} \left(1+ \frac{m_1^2-m_2^2}{P^2}\right)
\nnel
c^2 &=& \frac{E^2}{P^2} -
\frac{E^2 }{4m^2_1}\left(1+ \frac{m_1^2-m_2^2}{P^2}\right)^2
\nonumber
\eea
Comparing this result against the high temperature results already
obtained independent of \tref{ebhit} (only \tref{ebqhit}
was derived directly from \tref{ebhit}) we see that there is complete
agreement.

It is therefore clear that both time-ordered and retarded functions
are not analytic near zero-four momenta when the masses are equal.
In particular whenever the zero energy  limit is last, after or
taken with the $m_1=m_2, p$ limits, various answers can be obtained.
 If the masses are kept different \tcite{TSEbub,AVBD} or
three-momentum non-zero till last then a unique answer is obtained.
 The results are summarised in figure
\tref{fbcres}.  \if\tsepaper\tsefalse
\typeout{figure: BUBble RESults}
\begin{figure}[htb]
\tseepsfxsize{6in}
\tseepsffile{bubres.eps}
\caption{\tftitletwo}
\label{fbcres}
\end{figure}
\fi Starting at
the centre the energy integral or sum has been performed.  The
limits are taken in the order indicated as one moves out along a
radius.  The result obtained for $B_c$ is shown at the end of a
radius.  In figure \tref{fbcres}
it is given as the integrand for a remaining $\omega$
integral where a factor of $-g^2/8\pi^2$ has also been taken out
c.f. \tref{ebvinf},\tref{ebv},\tref{ebq}.  The high temperature
equivalents are given in figure \tref{fbcresht}
c.f. \tref{ebvinfhit},\tref{ebvhit},\tref{ebqhit}.
\if\tsepaper\tsefalse
\typeout{figure: BUBble RESults at High Temperature}
\begin{figure}[htb]
\tseepsfxsize{6in}
\tseepsffile{bubresht.eps}
\caption{\tftitlethree}
\label{fbcresht}
\end{figure}
\fi
Either RTF or ITF can be
used to calculate these results.

The answer obtained when the $E\ra 0 $ limit is not taken last is of
particular note because it is also the answer obtained if no
analytic continuation is done in ITF, where $E,p=0$ and $m_1=m_2$
are all set at the very start before any integration is done.  In
this case the order of the limits is irrelevant because the
integrand is analytic.  It is not clear why these limits coincide
but it means one can find this result by keeping $m_1\neq m_2$ or
$p\neq 0$ till the end.  The results obtained in these cases
are the same as the one used in free energy calculations, \tref{eep}.

The real part of the retarded function is
always finite as $E,p \ra 0$.  However the imaginary part is not always
zero its appearance reflecting the lack of analyticity despite the
fact that it is linear in the energy, \tref{eimc} and \tcite{TSEwpg}.
Thus there are peculiarities hiding in
this limit which are not obvious from the ITF formalism usually used
to calculate the retarded functions.

The time-ordered function, $B^{11}$ of RTF, is found to have  the
same real part as the retarded and so it is also finite.
However the imaginary part shows the effect of the problematic
$\delta^2$ term of RTF which results in the same $|\vecp|^{-1}$
behaviour as is found in the retarded function as $E,p \ra 0$
if $E<p$ and provided $m_1=m_2$.  Thus
the $\delta^2$ reflects a genuine divergence in the time-ordered
function.  It is, however, zero and well behaved when  $m_1\neq m_2$
or $p \neq 0$.

The problems with the bubble diagram and its various limits may seem
to be mere technicalities but they do have real do have real
implications for calculations of physical quantities.  One  example
of interest is the use of zero four-momentum diagrams to calculate
approximations to the finite temperature effective action
\tcite{TSEwpg}.  A detailed discussion given in \tcite{EEV}.
This might be done in order to obtain some dynamical information
such as is used in studying the electroweak phase transition (see
\tcite{pt} and references therein).    Usually a derivative
expansion on the effective action  is being performed.  At zero
temperature the lowest order term is just the effective potential
and such an expansion is well defined \tcite{Raybook}.  The
effective potential is thus generated by the zero-momentum 1PI
diagrams.  The various results for the zero four-momentum limit of
diagrams at finite temperature therefore calls into question the
simple derivative expansion of thermal effective actions but only
when there are equal masses or suitable self-interactions in the
theory.  This means that the coefficients of the expansion depend on
the order in which the time and space derivative expansions are
taken.   Thus, the results given here suggest that the lowest order
term is only necessarily the standard free energy if the time
derivative expansion was done first i.e. a static limit taken, $E\ra
0$ first.  This does not seem to be very relevant to dynamics where
one might expect a homogeneous, $p\ra0$ limit to be more relevant
but the example of the bubble diagram suggests that the lowest order
term is quite different in this case. The calculations performed
here and in \tcite{TSEbub,AVBD} stress the role of equal masses in
this problem of zero four-momentum limits.

Finally note that RTF and ITF give completely consistent results for
the bubble diagram (c.f. comments in \tcite{BD,We3,Xu}).  Either can
be used to calculate any of the functions discussed here.  While the
RTF has obvious singularities in the zero four-momentum equal mass
limit, the $[\delta(K^2-m^2)]^{N\geq 2}$, the same singularities are
hiding in ITF calculations of time-ordered functions.  Related
problems appear in calculations of the retarded
function in the unexpected non-zero, if finite, imaginary parts.

\if\tsenoteon\tsetrue { \section*{Appendix}

\bea
\int_0^\infty dk \; \frac{k^2}{\omega} n(\omega) &=& m^2 F_1(\beta m)
\nnel
&\simeq & \frac{\pi^2 T^2}{6} - \frac{T m \pi}{2 } ??? + ...
\nnel
\int_m^\infty d\omega \; \frac{1}{k} n(\omega) &=& F_2(\beta m) =
- \frac{1}{m} \frac{d}{dm} m^2 F_1(\beta m)
\nnel
&\simeq &\frac{\pi}{2} \frac{T}{m}
\eea

\bea
I_1 &=& \int_1^\infty dx \; n(b x) \nnel
&\simeq & \frac{1}{b} ( \zeta_0(1) + \psi(1) - \log (b) )
\tselea{eI1hit}
\\
I_2 &=& \int_1^\infty dx \; n^2(b x)
\nnel
&=& -I_1 + \frac{n(b)}{b}
\nnel
&\simeq & \frac{1}{b^2} +
\frac{1}{b} ( -\half -\zeta_0(1) - \psi(1) +\log (b) )
\tselea{eI2hit}
\eea
as noted in `RTF bubble Imaginary parts' notes, pp2, (27-9-92).
}
\fi

\npagepub
\section*{Acknowledgements}

I would like to thank the Royal Society for their support through a
University Research Fellowship.

\npagepub

\typeout{--- references ---}

\if\tsepaper\tsetrue \npagepub

\typeout{--- Hand made list of figures ---}

\section*{List of Figures}
\begin{description}
\item[\tref{fbub}] \tftitleone
\item[\tref{fbcres}] \tftitletwo
\item[\tref{fbcres}] \tftitlethree
\end{description}
\npagepub
\fi

\if\tsepaper\tsetrue
\typeout{figure: BUBble diagram}
\begin{figure}[thb]
\tseepsffile{bub.eps}
\caption{\tftitleone}
\label{fbub}
\end{figure}
\fi

\if\tsepaper\tsetrue
\typeout{figure: BUBble RESults}
\begin{figure}[thb]
\tseepsfxsize{6in}
\tseepsffile{bubres.eps}
\caption{\tftitletwo}
\label{fbcres}
\end{figure}
\fi

\if\tsepaper\tsetrue
\typeout{figure: BUBble RESults at High Temperature}
\begin{figure}[thb]
\tseepsfxsize{6in}
\tseepsffile{bubresht.eps}
\caption{\tftitlethree}
\label{fbcres}
\end{figure}
\fi

\end{document}